\documentclass[prl,preprint,superscriptaddress]{revtex4}

\usepackage{graphicx,bm}
\usepackage{color}

\begin{document}

\title{Water structuring and collagen adsorption at \\ hydrophilic and hydrophobic silicon surfaces}

\author{Daniel J. Cole}
\email{djc56@cam.ac.uk}
\affiliation{Theory of Condensed Matter Group, Cavendish Laboratory, \\
           University of Cambridge, J J Thomson Avenue, Cambridge CB3 0HE, UK}
\author{Mike C. Payne}
\affiliation{Theory of Condensed Matter Group, Cavendish Laboratory, \\
           University of Cambridge, J J Thomson Avenue, Cambridge CB3 0HE, UK}
\author{Lucio Colombi Ciacchi}
\email{colombi@hmi.uni-bremen.de}
\affiliation{Hybrid Materials Interfaces Group,
             Faculty of Production Engineering \\ and Bremen Centre for Computational 
             Materials Science, University of Bremen, 28359 Bremen, Germany}
\affiliation{Fraunhofer Institut f\"ur Fertigungstechnik und Angewandte Materialforschung
             IFAM, Wiener Str. 12, 28359 Bremen, Germany}

\begin{abstract}

The adsorption of a collagen fragment on both a hydrophobic, hydrogen-terminated 
and a hydrophilic, natively oxidised Si surface is investigated using all-atom 
molecular dynamics.
While favourable direct protein-surface interactions via localised contact points 
characterise adhesion to the hydrophilic surface, evenly spread surface/molecule
contacts and stabilisation of the helical structure occurs upon adsorption on the 
hydrophobic surface.
In the latter case, we find that adhesion is accompanied by a mutual fit between 
the hydrophilic/hydrophobic pattern within the protein and the layered water structure 
at the solid/liquid interface, which may provide an additional driving force to the 
classic hydrophobic effect.

\end{abstract}

\maketitle

The issue of protein adsorption onto solid surfaces in a wet environment is of great 
importance for the design and functionalisation of materials in the contexts of
biomedical implants, biosensors, antifouling surfaces, pharmaceutical packaging,
and many others~\cite{biomems,immobilisation,sci-art}.
One important case is the adsorption of extracellular matrix proteins
such as collagen on the surfaces of implanted Si-based
microelectromechanical systems~\cite{biocom-si}.
Control over protein adhesion may be achieved by manipulation of surface 
properties, such as isoelectric point, functional group termination, 
topography and hydrophobicity~%
\cite{physiol,jusu,keselowsky,wilson-fn,kimkim,garciaphob,jonsson-explains,vanwachem}.
However, the atomistic details of the interactions mediated by these
effects are still unclear and are the object of intense experimental and
theoretical investigation~\cite{pnas-phobic,raff-graph,raff-review,biomat}.

In a previous work~\cite{wafer_paper}, we have shown that liquid 
water close to a natively oxidised Si surface presents alternating layers 
of larger and smaller density with respect to the bulk.
These structural features directly influence the strength of mutual adhesion 
between pairs of Si wafers at room temperature under humid conditions~%
\cite{wafer_paper}.
In a similar way, the adsorption and adhesion of biomolecules to
solid surfaces may be significantly determined by the structure of 
water layers in the surface proximity~\cite{hower,bujnow,cormack,zangi}.
In particular, the disruption of the adsorbed water layer by individual amino 
acids has been shown to have both enthalpic and entropic effects on 
the free energy of the bound system~\cite{latour2002}.
In this work we show through all-atom molecular dynamics simulations 
(MD)~\cite{coll_meth} that a complex oligopeptide interacts with a 
Si surface via a mutual fit between its intrinsic hydrophilic/hydrophobic
pattern and the structured water layers close to the surface.
This may provide a driving force for adsorption at the solid/liquid
interface which acts in addition to the classic hydrophobic 
effect~\cite{pnas-phobic,vogler98}.

Specifically, we use MD to investigate the adsorption behaviour of the NC1(84-116) 
domain of collagen XIV (PDBID:~1b9q), on both a hydrophilic (i.e.\ natively
oxidised~\cite{Lucio-water}) and a hydrophobic (i.e.\ hydrogen-terminated) 
Si(100) surface.
The $\alpha$-helical protein conformation used as the starting point for our simulations 
has been found by circular dichroism measurements to be stabilised by interaction with 
hydrophobic surfaces~\cite{coll-func}.
The folded collagen structure presents one predominantly hydrophobic side and a series 
of hydrophilic basic residues (lysine) clustered on the opposite side.
These basic residues form a specific binding site through which 
collagen XIV interacts with heparin molecules~\cite{coll-func}.
Here, we consider adsorption of the $\alpha$-helix on solid surfaces with 
the heparin binding lysine residues pointing towards the solvent, away from 
the surface.

In our simulations, water molecules are treated using the TIP3P 
force field~\cite{tip3p} and protein-water and protein-protein 
interactions are described by the {\textsc amber} parm99 biomolecular 
force field~\cite{amber10}.
The water solvent and the protein interact with the surface via Coulomb 
and Lennard-Jones (LJ) interactions.
The atomic charges and interatomic interactions for the natively 
oxidised Si surface are taken from Ref.~\cite{wafer_paper}.
The LJ energies $\epsilon_i$ and radii $\sigma_i$ of the Si and O
surface atoms are accurately tuned to reproduce the correct heat of
immersion of hydrated silica surfaces~\cite{wafer_paper}.
In the case of the hydrophobic H-terminated Si(100) surface, terminal H 
atoms are assigned small ESP charges of -0.08~$e$  (as computed by 
density functional theory), which are compensated by charges of +0.08~$e$ 
on their neighbouring Si atoms.
For the H atoms of the hydrophobic surface, we use the same LJ
parameters as for hydrophobic H atoms of hydrocarbon chains.
The surface-protein LJ interactions are obtained by combining the
surface and biomolecular force field parameters using the 
standard Lorentz-Berthelot combination rules~\cite{LJcombinationrule}.
We have performed both shorter simulations allowing all atoms of the surfaces to
move and longer simulations where the surfaces remain frozen throughout, except 
for hydroxyl H atoms on the natively oxidised surface, which are free to rotate 
as described in Ref.~\cite{wafer_paper}.
The water structure at the rigid natively oxidised surface is nearly identical to 
that obtained for the unconstrained surface, as also reported in~\cite{wafer_paper}.

Two simulations are started initially placing the collagen helix a few \AA\ above 
two randomly chosen sites of the natively oxidised Si surface~\cite{coll_meth}.
Close contact between the surface and collagen is achieved by performing 
a 2000 step conjugate gradients minimisation in vacuum.
The entire simulation cells are then filled with water molecules and the cell 
heights are adjusted to obtain the equilibrium TIP3P water density far from the surfaces.
The solvated systems are slowly heated to 300~K applying weak harmonic 
constraints to the protein and, finally, constant-temperature production 
MD runs are performed for 24~ns.

The collagen fragment is found to be highly mobile on the hydrophilic surface 
(Fig.~\ref{figa}~(top)).
In the first simulation, the molecular centre of mass initially moves away 
from the surface, while localised contact (within 3.4~\AA{} of the surface) 
is maintained via residues 108--112, resulting in an end-on bound conformation.
Later a second point of contact via residues 91--93 is formed.
The position of the centre of mass remains stable at approximately 9~\AA{} 
over the final 10~ns of the simulation, and the root mean square 
displacement of the backbone atoms of the protein converges to within 1~\AA{}.
In the second simulation, two points of contact are maintained throughout most 
of the 24~ns simulation, again via the molecular termini (roughly localised in
the regions of residues 84--88 and 104--112).
Interestingly, residues Leu107 and Arg111 remain in close contact with the surface
throughout both simulations via the {\em hydrophobic} CH$_n$ part of their side 
chain groups.
The total energy of adsorption averaged over the final 10~ns of the two simulations is 
$-22\pm7$~kcal/mol with respect to a reference unbound simulation in which collagen is 
separated from the surface slab by at least 10~\AA{}.

A different behaviour is observed in two similar simulations (each lasting 14~ns) 
performed starting with collagen in the same conformation over two different 
starting positions close to the hydrophobic, H-terminated Si surface.
The fragment remains adsorbed to the surface and maintains its helical structure 
throughout both simulations (Fig.~\ref{figa}~(centre)), again with an average  
energy of adsorption of $-22\pm3$~kcal/mol with respect to a corresponding 
unbound system.
Contrary to the case of the hydrophilic surface, the residues in direct contact with 
the surface are evenly spread through the central part of the protein (residues 
91--108), while the molecular termini (84--90 and 109--116) are less tightly bound.

As visible in the Ramachandran plot computed over the final 10~ns of each simulation (Fig.~\ref{figb}), 
the backbone dihedral angles of collagen on the hydrophobic surface are tightly clustered around the ideal 
$\alpha$-helix values (-65$^\circ$,-35$^\circ$).
A time-averaged calculation of the $\alpha$-helix content of the protein using the DSSP 
algorithm~\cite{DSSP} reveals that $33~\%$ of the residues have an $\alpha$-helical structure.
This is mainly concentrated in the central protein region of residues 93--112, in qualitative agreement 
with a structural assignment based on NMR spectroscopy~\cite{coll-func}.
For comparison, in a separate 14~ns simulation of collagen in bulk water, there is a much wider spread of 
dihedral angles, with an increase of their occurrence in the $\beta$-sheet region.
With respect to the adsorbed case, the average helical content falls to $18~\%$, and the average number of 
intramolecular hydrogen bonds decreases from 19 to 17 (as computed with a distance cut-off of 3.4~\AA{} 
and an angular cut-off of 140$^\circ$).
On the hydrophilic surface, the degree of unfolding is intermediate between the hydrophobic 
and unbound structures (Fig.~\ref{figb}) and 20 internal hydrogen bonds are maintained 
on average.

Although the degree of $\alpha$-helical conformation in bulk water may be over-estimated
due to our helical starting structure, the uncoiling observed in the time scale of these 
simulations appears to be significant.
In Ref.~\cite{coll-func}, it is contended that unfavourable electrostatic interactions 
between arginyl and lysyl side chains in the region of the heparin binding site 
contribute to helix destabilisation which causes the collagen domain to unfold in water.
Indeed, by computing the pairwise electrostatic interactions involving only the primary heparin 
binding region (residues 97-105), we find that the net electrostatic energy is approximately 
30~kcal/mol less favourable at the hydrophobic surface than in bulk water,
where the Coulomb repulsion between charged residues is lowered as a consequence 
of the helical unfolding.

These results are consistent with the experimental observations that
the $\alpha$-helical form of the NC1 collagen XIV fragment adsorbs stably 
on hydrophobic organic surfaces~\cite{coll-func}.
Here we observe that collagen remains adsorbed at the hydrophobic Si-H surface for a 
total of 28~ns with a favourable net binding energy.
On the one hand, displacements of water molecules close to non-polar 
amino acids following direct interactions with the surface, together with the 
associated re-arrangement of the solvent hydrogen-bond network, 
are indeed known to drive adsorption and represent a fundamental part
of the so-called hydrophobic attraction~\cite{pnas-phobic,vogler98}.
On the other hand, more subtle structural effects arising from the presence of 
an abrupt solid/liquid interface combined with a favourable arrangement 
of the adsorbed protein in an ordered hydrophobic/hydrophilic pattern
may also contribute to stabilize the adsorption, as we shall see below.

The effects of the molecular water structure on the biomolecular adsorption are
studied by analysing the density profiles of water and protein H atoms along the 
direction perpendicular to the surface.
In the case of water, this is computed by counting the 
average number of H atoms present in thin planar slices of width 0.1~\AA{} in 
separate 1~ns simulations in the absence of collagen.
The disruption of the bulk water hydrogen bond network by the two 
surfaces results in a re-ordering of interfacial water molecules to 
maximise their interactions with neighbouring molecules and 
the surface.
This is manifested by the structuring of water into layers 
of higher and lower density with respect to the bulk, as found previously 
in numerous MD simulations~\cite{wafer_paper,rossky_phob,michaelides08},

In the case of the hydrophobic surface (Fig.~\ref{figc}(a), black line), the
first peak in the water density is separated from the surface H atoms (grey lines) 
by 2.7~\AA{}, consistent with weak surface-water interactions of the 
Lennard-Jones type~\cite{heat_imm_hydroph}.
Two other maxima are visible at 5.9 and 9.1~\AA{} far from the surface.
This surface-induced water layering is preserved in the presence of 
adsorbed collagen, as evident from the density profile of the 
150 water molecules belonging to the first two solvation shells of the 
protein fragment (dotted line).
The first three peaks are centred at 2.8, 6.5, and 9.7~\AA{}, only
slightly shifted from those observed in the absence of the protein. 
We now compute in the same way the density of H atoms belonging to collagen, 
averaged during the MD trajectories described above, discriminating hydrophobic 
H atoms (bonded to C atoms) from hydrophilic H atoms (bonded to O, N or S atoms).
The H atoms of the protein are clearly structured in alternating regions of 
higher and lower density, revealing an intrinsic ordering of hydrophobic/hydrophilic 
residues in the adsorbed helical structure (Fig.~\ref{figc}(a)).
Interestingly, the hydrophobic/hydrophilic pattern within the adsorbed 
protein is commensurate with the water layering oscillations in the surface
proximity.
Namely, the first water peak is located between the first and second 
small peaks of the hydrophobic H atoms, centred at 2.2 and 3.9~\AA{}.
The latter peak coincides approximately with the first trough in the water 
density (located at 4.4~\AA{}).
Correspondingly, the large trough in the hydrophobic H density at 6.2~\AA{}
coincides with the second water density peak.
Further, the first peak in the hydrophilic H density is located at 
2.8~\AA{}, in perfect correspondence with the large first water peak.
The broad main peak overlaps with the second water peak and the third
peak (corresponding to the cluster of lysine residues) points away 
from the surface/protein interface into the bulk solution~\cite{unconstrained}.

In the case of the natively oxidised surface, water molecules penetrate 
into the oxide layer and the main density peak is centred on the surface 
hydroxyl groups (Fig.~\ref{figc}(b)).
In contrast to the clear structuring observed at the hydrophobic surface, both 
hydrophilic and hydrophobic H atoms of the protein show relatively spread and less 
structured density profiles, and the matching between the hydrophilic/hydrophobic
patterns in the molecule and the water layer structure is barely visible. 
Surprisingly, while hydrophilic H atoms of the protein are shielded from the surface 
by the first high density water peak, the H atoms of hydrophobic groups are the closest 
to the surface.
This is consistent with our previous finding that predominantly hydrophobic side 
chain groups remain in close surface contact (see above), bound via van der Waals 
forces to bridging oxygen atoms of the oxide network, which are hydrophobic compared 
to the surface silanol groups~\cite{wafer_paper}.

Despite the evident ordering of water molecules at the surface/protein interfaces,
their mobility within the protein hydration shells remains of the same order 
of magnitude before and after adsorption.
This is shown by the mean residence times of the first shell water molecules surrounding 
the wet surfaces, the adsorbed proteins and the protein in bulk water, calculated with 
the method introduced in Ref.~\cite{mrt-water}
(Table~\ref{figd}).
As expected, the residence times of water molecules in the surface proximity, 
especially for the hydrophilic case, are higher than the value of 7~ps obtained 
for bulk water.
However, the residence times of water molecules around collagen (80~ps
for the dissolved protein, as found for similar systems~\cite{mrt-water}) 
are only slightly increased when the protein is adsorbed at the two surfaces.

To summarise, collagen remains bound to the oxidised Si surface via a few 
terminal residues, the most stable direct interactions with the surface 
being between hydrophobic side chain groups and hydrophobic surface 
sites.
Individual hydrophilic residues penetrate into the first water layer to 
form direct hydrogen bonds with the surface for only 20--30~\%\ of the 
total simulation time.
This is consistent with a suggested adsorption model in which proteins 
weakly bound to hydrophilic surfaces leave the surface-bound hydration 
layer almost intact and do not affect the surface tension at the solid/liquid
interface~\cite{vogler95,vogler98}.
In contrast, proteins tend to adsorb with larger contact surface areas 
to hydrophobic surfaces~\cite{physiol,jonsson-explains,vanwachem}.
In particular, the NC1 domain of collagen XIV has been observed to
adsorb stably in an $\alpha$-helical structure on a hydrophobic 
organic layer~\cite{coll-func}, as supported by our simulations on
the H-terminated Si surface.

We note that an accurate analysis of the individual enthalpy and entropy contributions 
leading to the observed adsorption behaviour is made difficult by the large oscillation 
of each separate energy term and the intermittent exchange of counter-ions between surface, 
protein and bulk water.
Extensions of this work will be necessary to calculate the full potential of mean force 
along possible adsorption/desorption paths at different temperatures~\cite{pnas-phobic}.
In our analysis so far, we have focused on the structural details governing 
collagen adsorption on silicon and provided evidence for a strict interplay 
between the solvent structure at the solid/liquid interface and the intrinsic 
hydrophobic/hydrophilic pattern within the adsorbate.
Whether this effect is specific to systems that are intrinsically
chemically commensurate with the water layering, or whether flexible
biopolymers may indeed adjust their internal structure to fit the
solvent ordering at the solid/liquid interface (in a process
reminiscent of the ``induced fit'' mechanism for supramolecular
organisation~\cite{Magali_2007}) is a question that will require
further combined theoretical and experimental efforts.

\begin{acknowledgments}

Computer resources were provided by the Cambridge High Performance 
Computing Service, UK, and by the Center for High Performance Computing 
(ZIH) at the University of Dresden, Germany.
We acknowledge support by the EPSRC, the Deutschen Forschungsgemeinschaft 
within the Emmy-Noether Programme (CI 144/2-1), and the European Community under the 
HPC-EUROPA project (RII3-CT-2003-506079) and the EU-FP7-NMP Grant 229205 ``ADGLASS''.

\end{acknowledgments}

\clearpage

\subsection*{Figure Captions}

\begin{figure}[h!]
\caption{MD trajectories of collagen on a natively oxidised (top) and a 
H-terminated (centre) Si(100) surface and in a separate simulation in bulk water, 
in the absence of any surface (bottom). Initial (left), intermediate (centre) and 
final (right) snapshots are pictured. Collagen consists of 34 amino acid residues 
and its predominantly hydrophobic surface is initially oriented towards the 
surfaces. Water molecules are omitted for clarity.
}
\label{figa}
\end{figure}

\begin{figure}[h!]
\caption{Ramachandran plots showing the distributions of the dihedral angles, 
$\phi$ and $\psi$ (degrees), for collagen at the hydrophobic surface (left), 
hydrophilic surface (centre) and in bulk water (right).}
\label{figb}
\end{figure}

\begin{figure}[h!]
\caption{Density profiles of hydrophobic (green) and hydrophilic (blue) 
H atoms of collagen at (a) a H-terminated and (b) a natively oxidised Si surface. 
Density profiles of water H atoms (black, solid line) are obtained from separate 
1~ns simulations. Also shown are the H atoms of water molecules belonging to the 
first two solvation shells of collagen (black, dotted line) and H atoms of the 
surfaces (grey).
}
\label{figc}
\end{figure}

\begin{figure}[h!]
\caption{Water mean residence times (ps) about the protein adsorbed at the hydrophobic 
and hydrophilic surfaces and about the protein in bulk water. Also shown are the residence 
times of water molecules at the two surfaces in the absence of collagen. The two values quoted at 
hydrophilic surfaces are for hydroxyl groups and bridging oxygen atoms respectively. Snapshots are 
analysed every 2~ps in sliding windows of 400~ps in the final 10~ns of each simulation. 
The mean residence times are obtained by fitting a stretched exponential to the autocorrelation 
functions of the boolean variable that indicates whether or not a given water molecule 
is within a shell of width 3.5~\AA{} around the protein or surface.
}
\label{figd}
\end{figure}

\clearpage

\begin{center}
\includegraphics[width=0.9\textwidth]{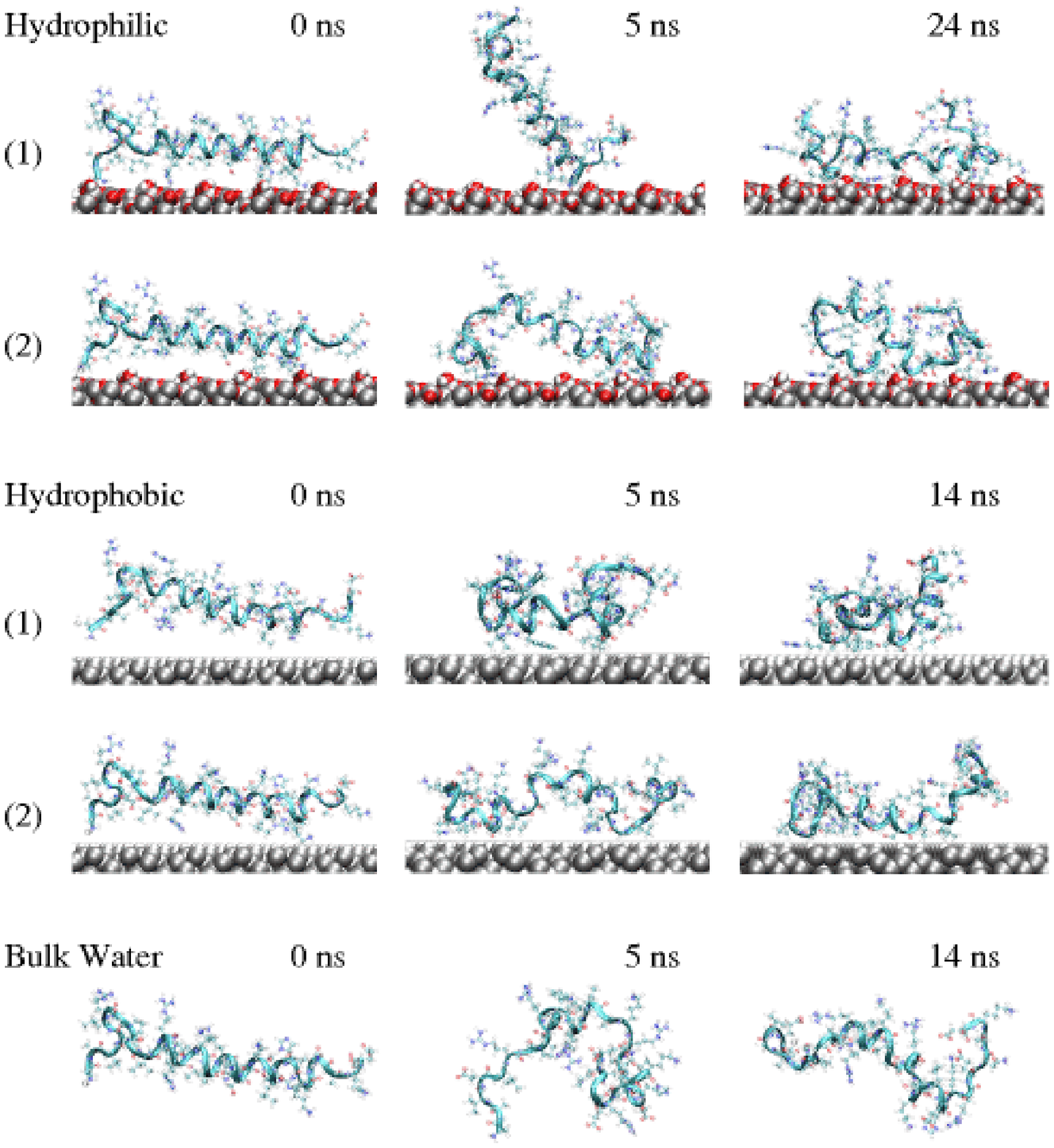}
\end{center}
\vfill\
Figure~\ref{figa}, D. J. Cole et al.

\clearpage

\begin{center}
\includegraphics[width=0.9\textwidth]{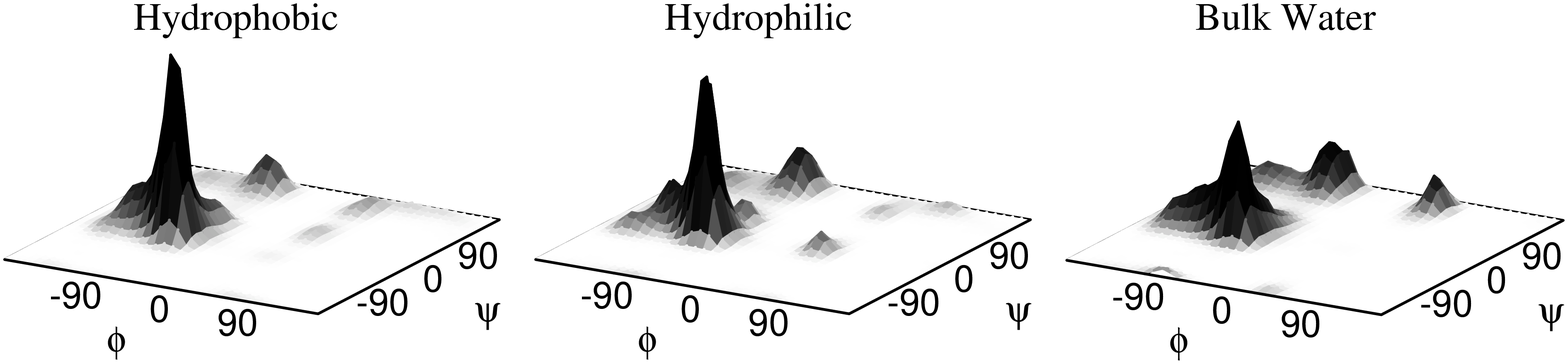}
\end{center}
\vfill\
Figure~\ref{figb}, D. J. Cole et al.

\clearpage

\begin{center}
\includegraphics[width=0.9\textwidth]{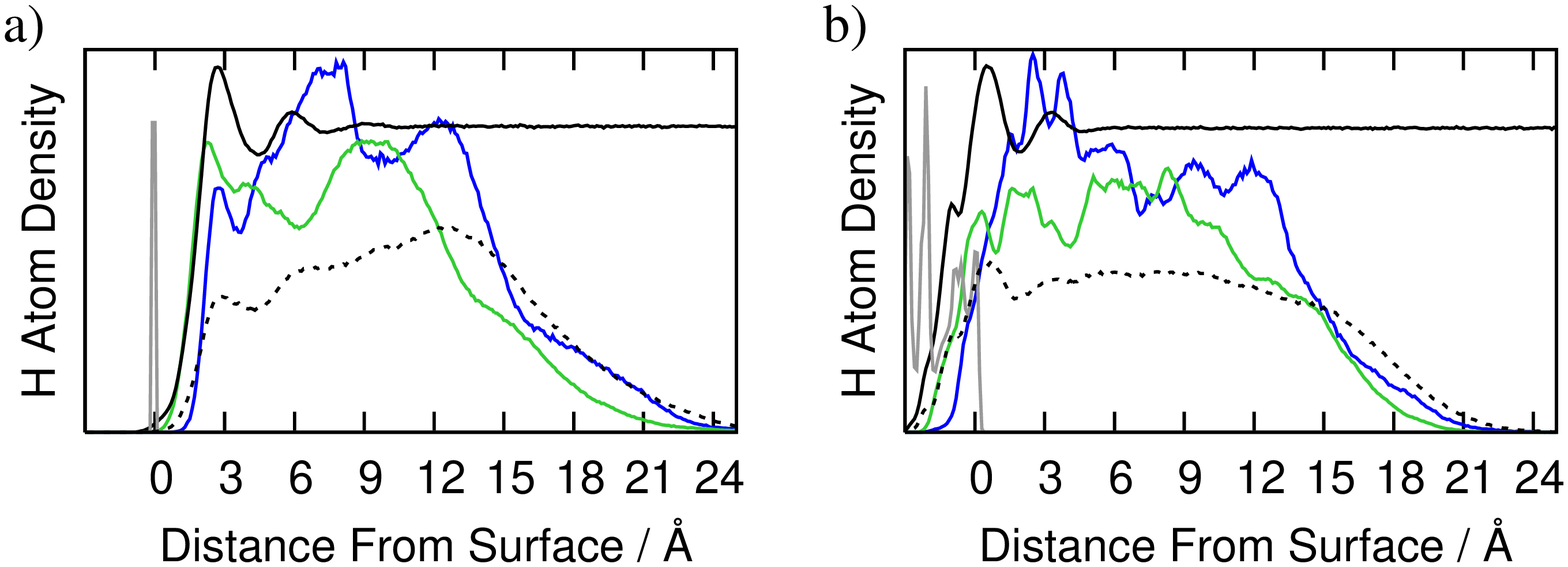}
\end{center}
\vfill\
Figure~\ref{figc}, D. J. Cole et al.

\clearpage

\begin{table*}[ht!]
\begin{center}
\begin{tabular}{lccccc}
 & \multicolumn{2}{c}{\hspace*{0.5cm}Hydrophobic\hspace*{0.5cm}}  &  
   \multicolumn{2}{c}{\hspace*{0.5cm}Hydrophilic\hspace*{0.5cm}}  & 
   \hspace*{0.5cm}Bulk\hspace*{0.5cm} \\
 &\hspace*{0.5cm} (1)    &   (2)   &  (1)    &   (2)  &      \\[1mm]
\hline \hline
Protein\vphantom{\Large A}  & \hspace*{0.5cm} 98  &  84  & 118      &  126     & 80 \\[1mm]
Surface                     & \hspace*{0.5cm} 37  &  36  & 157/214  &  149/200 & -- \\[1mm]
\hline
\end{tabular}
\end{center}
\end{table*}

\vfill\
Figure~\ref{figd}, D. J. Cole et al.


\begin{thebibliography}{37}
\bibitem[{col()}]{coll_meth}
\bibinfo{note}{Our MD simulations are performed with the
  \textsc{amber10}~\cite{amber10} package, using an integration time
  step of 2~fs and the SHAKE algorithm to constrain H-containing bonds.
  Coulomb interactions are treated using the Particle Mesh Ewald sum, with a real 
  space cut-off of 10~\AA{}. The cut-off length for LJ interactions is also set to
  10~\AA{}. The temperature is controlled by the weak-coupling algorithm with a 2~ps time
  constant and, during equilibration, the systems are heated at a rate of 50~K per 
  100~ps. The protein fragments are terminated with --NH$_3^+$ and --COO$^-$
  groups, and seven Cl$^-$ ions are added to ensure charge neutrality. Si(100)
  surface models of area 95.7~nm$^2$ are obtained by repeating a ($2\!\times\!2$)
  unit cell. The surface slabs are placed in a supercell subject to periodic boundary 
  conditions and separated by a vacuum gap of approximately 60~\AA{}. The Si lattice 
  parameter is fixed to the equilibrium value obtained with the Stillinger-Weber potential 
  (5.44~\AA) and the oxide structure is taken from Ref.~\cite{Lucio-water}.
  Structural and energetic properties are calculated over the final 10~ns in each case.
  For the non-rigid surface simulations, we use the surface force field described 
  in~\cite{wafer_paper} and the \textsc{dl$\_$poly$\_$3} MD package.
 }
\bibitem[{hea()}]{heat_imm_hydroph}
\bibinfo{note}{The calculated heat of immersion of the model H-terminated Si
  surface, $-22$~mJ/m$^2$, is in good agreement with values inferred from
  contact angle measurements of H-terminated gold
  surfaces~\cite{phob-himm-result}.}
\bibitem{unconstrained}
Notably, the same ordering between the water and collagen structure is
observed in shorter simulations ($\sim$2~ns) in which all atoms of the
surface are allowed to move.
\end{thebibliography}
\end{document}